\documentclass[]{aa}
\usepackage{graphics}

\newcommand \mic{${\,\mu {\rm m}}$}

\newcommand \ad {ADONIS}
\newcommand \ngc {{\sc ngc\,1068}}
\newcommand \ooo {[O\,III]}
\newcommand \ap{$\sim$}
\newcommand \dg {$^\circ$}
\newcommand \x {\times}

\begin{document}

\thesaurus{3(11.09.1 NGC 1068; 11.19.1; 11.14.1; 13.09.1; 11.01.2; 03.09.7)}
\title{Adaptive Optics Images at 3.5 and 4.8 \mic\
of the Core Arcsec of NGC\,1068:\\
More Evidence for a Dusty/Molecular Torus.
\thanks{Based on observations collected at the European Southern Observatory, La 
Silla, Chile.}}
\author{ O. Marco \and D. Alloin}
\institute{European Southern Observatory (ESO), Casilla 19001, Santiago 19, 
Chile}
\offprints{O. Marco}
\mail{omarco@eso.org}
\date{Submitted to Astronomy \& Astrophysics}
\titlerunning{Dusty/Molecular Torus in NGC\,1068}
\maketitle

\begin{abstract}
 
Adaptive optics observations of \ngc\ 
have allowed to reach with a 4\,m class telescope, 
diffraction-limited images at 3.5 \mic\ (FWHM=0.24'')
and 4.8 \mic\ (FWHM=0.33'') of the
central arcsec region.
These observations reveal the presence of an unresolved core 
(radius less than 8\,pc at Half Maximum) and
an 80\,pc sized ``disc-like'' structure at P.A. \ap 100\dg,
which is interpreted as the dusty/molecular torus invoked in the
AGN unification scheme. They show as well an extended emission region 
along the NNE-SSW direction, 
of 100\,pc full size, most probably associated with dust in the NLR.
The position of the unresolved core at 3.5 and 4.8 \mic\ 
is found to be coincident with that of the core observed 
at 2.2 \mic\ and outlines the location of the central engine
in the AGN of \ngc.
Considering as well previous AO observations at 2.2 \mic\ 
we infer that there must exist a very steep gradient of the dust
grain temperature, close to the central engine. At 
a distance of 30\,pc from the central heating source, the dust grain 
temperature deduced from the [L-M] color is found to be \ap 500\,K.
The mass of warm dust ($T_{gr} >$500\,K) in the
0.6'' diameter core
is found to be \ap 0.5~$M_{\odot}$.
The spectral energy distribution from 1 to 10 \mic\ is provided for 
the 0.6'' diameter core.
These results are briefly discussed in the context of current torus 
models.


\end{abstract}

\section{Introduction}

Series of observational facts assembled over the past decade on Active
Galactic Nuclei have led to the so-called ``unified'' model of AGN
(for a review of these facts, as well as for the detailed
characteristics of the unified model see Krolik 1999).
Our current specific interest in the unified model is that the central
engine (black hole and accretion disk) and its close environment
(dense gas clouds emitting the broad lines which constitute the broad
line region, BLR) are embedded within an optically thick dusty/molecular
torus. Along some lines of sight, the torus obscures and even fully
hides the central engine and the BLR.

In that respect, the case of \ngc, a bright Seyfert 2 active galaxy, 
 is particularly enlightening. The
spectrophotometry in polarized light reveals the presence of a hidden
BLR (Antonucci \& Miller 1985),
the conical shape of the narrow line region (NLR) -- both on large
(Pogge 1988) and small (Evans et al. 1991) scales -- indicates that the 
ionizing
radiation is collimated by an opaque blocking torus and, finally, 
the symmetry center of the UV/optical polarization map 
(Capetti et al. 1995) is found to be coincident
with the radio core (Gallimore et al. 1996a), the 12.4 \mic\ peak
(Braatz et al. 1993) and the maser emission
(Gallimore et al. 1996b), suggesting that this is
the location of the hidden true nucleus.

This object appears then particularly suitable for unveiling the putative 
torus through its infrared emission, which, according to current
models should be quite strong ({\it e.g.} Pier \& Krolik 1992a, b, 1993; 
Granato \& Danese 
1994;
Efstathiou, Hough \& Young 1995;  Granato, Danese \& Franceschini 1997). 
In this search, high spatial resolution is a requisite
 in order to locate very precisely 
and to characterize the structure of emission sources. Hence,
adaptive optics (hereafter abbreviated AO) in the 1-5 \mic\ window is the 
tool. 
At the distance of \ngc\ (14.4\,Mpc), 1'' is equivalent to  72\,pc (assuming 
H$_0$=75 km/s/Mpc),
allowing to reach a spatial resolution of a few parsecs. 
Through AO observations, simultaneously in the visible and near-infrared, 
the 2.2\mic\ peak
emission has already been found 
to be offset by \ap 0.3'' S of the
optical continuum peak as defined by Lynds et al. (1991) and to be coincident with
the previously identified ``hidden true nucleus'' (Marco, Alloin \& Beuzit 1997). 
In the current
study, we are presenting new results obtained at 3.5 and 4.8\mic\
with ADONIS, the AO system working at the 
ESO 3.6\,m telescope on La Silla and fully described in Beuzit et al. 
(1994).

\section{Observations and data reduction}

The observing run took place from august 13 to 19, 1996,  under excellent
seeing and transparency conditions (Table \ref{data_ac}).

\begin{table}[t]
\caption{Summary of the data sets}
\begin{center}
\begin{tabular}{lcccccc}
\hline
filter	&$\lambda$	&T$_{int}$	&T$_{total}$	&Strehl	&visible 	&airmass	
\cr
	&[\mic ]	&[s]		&[s]		&[\%]	&seeing ['']			
\cr
\hline
L	&3.48		&6.0 		&1152 		&28 	&0.7 	&1.23	
\cr
M	&4.83		&1.5 		&720 		&36 	&0.55 	&1.15	
\cr
PAH	&3.31		&60 		&1440  		&25 	&0.66 	&1.25	
\cr
\hline
\end{tabular}
\end{center}
\label{data_ac}
\end{table}

The AO correction was performed on the brightest
spot of \ngc\ in the visible continuum light (Lynds et al. 1991).
 The wavefront
sensor (EBCCD after a red dichroic splitter) has a pixel size of
0.7\arcsec\ and takes into account the gravity center of the light
within a 6\arcsec\ diameter circular entrance. Due to the pixel size
of the wavefront sensor,
the contribution of the continuum and the lines from the central
50\,pc around the Lynds et al. peak (1991) both fall in one pixel: so the
contrast is maximum.

The detector was the COMIC 
camera (Lacombe et al. 1997), at the f/45 Cassegrain focus,  
which provides an image scale of 0.1\arcsec/pixel, resulting in a field
of view of 12.8\arcsec $\times$ 12.8\arcsec.

\ngc\ was observed in an imaging mode 
through the standard spectral L (3.48 \mic), L' (3.81 \mic), and M
(4.83 \mic) bands and through a
circular variable filter for the PAH (line 
3.3 \mic\ rest wavelength, 
and continuum). 
Through the L, L' and M bands, we observed in a chopping mode,
alternating object and sky images by the use of a field selection
mirror. We chose an offset of 10'' to the N and 10'' to the W. 

During the six nights, the visible seeing was measured by the ESO
differential image motion monitor. It was excellent during four nights, 
ranging from 0.4'' to 0.7''.
 Therefore, the efficiency of
the AO correction was optimal and the images obtained
with COMIC were diffraction-limited. However, the gain of the
intensifier was not optimized at that time, which resulted in Strehl ratios
 lower than expected.

In order to minimize position offsets between the calibration star and the
AGN, we selected a reference star within 2 degrees of the target.
For both the galaxy and the point spread function (PSF) reference star, the
air-mass was at most of 1.3, ensuring 
differential refraction effects to be negligible (less than one pixel).

Individual exposure times were chosen
so as to observe under conditions of background limiting performances
(BLIP). In this way, the readout-noise is dominated by the background
photon noise, and we just take an average of the images to improve the
signal-to-noise ratio. We observed several photometric standard stars to 
obtain a precise flux calibration 
and another star to determine the PSF for later deconvolution.

Standard infrared data reduction procedures were applied to each
individual frame, for both the galaxy and the reference stars : dead pixel
removal, sky subtraction, flat fielding from sky images at each wavelength. 
As the AO system is compensates for image
shifting, no additional shift correction was applied.

Thanks to the length of our observing run, we obtained a largely 
redundant data set. We discovered in particular that we had experienced
a problem of astigmatism during 2 nights, due to an unappropriate 
tuning of the visible wavefront sensor leading to a very low
 signal to noise ratio on the Shack-Hartmann analysor (Alloin \& Marco 1997).
Being aware of this flaw, we decided  to review in depth the AO
optimization parameters for the entire data set and to remove all
suspicious blocks. In addition, we applied
a selection procedure of  32-images data cubes,
based upon the seeing value and the Strehl ratio.
The corresponding equivalent integration times are reported in 
Table \ref{data_ac}.

\begin{figure*}
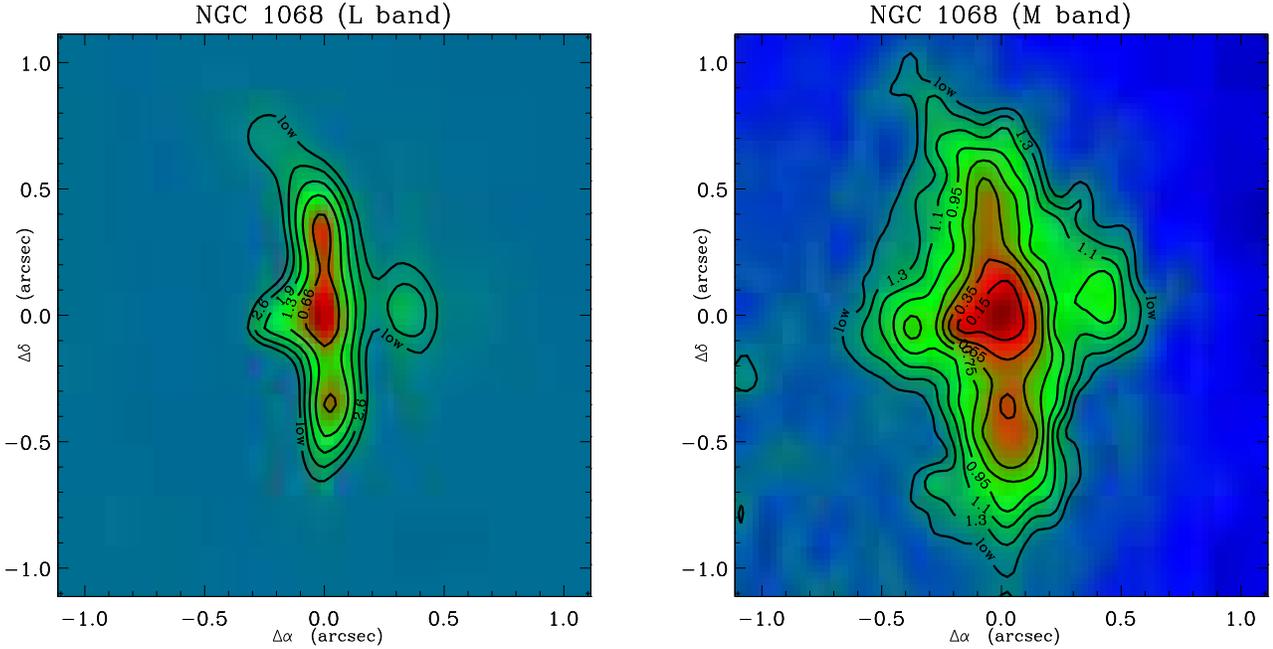

\resizebox{\hsize}{!}{\includegraphics{MS8699.f1a}\includegraphics{MS8699.f1b}}
\caption{Left: L band image of \ngc. Right: M band image of \ngc. North is up, 
East is to the left.}
\label {lmimage}
\end{figure*}

\section{Wavefront analysis sensing}

To better interpret the \ngc\ observations with \ad, we need
to track precisely which visible image of \ngc\ is seen by the 
wavefront analysis
sensor (WFAS) in its evaluation of the AO
correction. 

In order to recover this information, we have used HST images with
 45 mas pixel size both in the continuum and in the lines (F791W, F547M
F658N, and F502N), properly aligned and flux scaled to a same 1 sec
exposure at the entrance of the telescope, to construct a composite
 image which mimics the one seen by the WFAS.
 The image alignment was performed using point-like sources in the
field of view, which allows a registration within 10\,mas
(Tsetanov, private communication). The flux scaling has
included corrections for the exposure time of each image and the
 mean HST camera efficiency over the corresponding filter. Then, each
of these corrected images has been weighted by the corresponding mean  WFAS
wavelength response, before addition. We find  indeed that the light
received from \ngc\ on the WFAS is largely dominated by continuum light
and is not much weighted by the \ooo \,-\,line light distribution. 

And finally, to mimic the image seen on the WFAS CCD, from which the
AO corrections are computed, we have degraded the HST composite image by a
``seeing'' effect of \ap 1'': this includes both an atmospheric effect
(\ap 0.6'' according to the mean seeing value -FWHM- measured on the selected 
nights of the
observing run) and the instrumental spread function (0.85''), which is
rather large because of the photocathode -CCD spacing. 
As the  WFAS takes into account, to compute the AO
correction, the light centroid over a 6'' circular diaphragm, 
we have derived the light centroid within a 6''
aperture on the degraded composite HST image. This position is the
reference for the WFAS.
 
We need also to locate the centroid of the degraded HST composite image
with respect to 
the Lynds visible peak, as measured from the HST image F547M. 
We find a slight offset between the visible centroid and the Lynds visible
 peak: the visible centroid (using \ngc\ for the correction) is
located 99 mas to the N and 86 mas to the E of the Lynds
peak. This offset is taken into account in our estimate of
the \ngc\ infrared sources location (Sect. 4.1). The offset may 
differ from one AO system
to another. In the case of \ad, the offset is largely due to the 6'' entrance
to the WFAS.

\section{Data analysis}

We present in Fig. \ref{lmimage} the L and M band images
of \ngc. We used a magnitude (log) scale because of the high dynamics of 
the images
provided by the AO. The images have been deconvolved using a 
Lucy-Richardson algorithm (MIDAS package).
We have also observed \ngc\ in the L' band: this image is very similar 
to the one obtained  in the  L band.
Thus,  our current data analysis and subsequent
discussion  will be based on the two L and M bands, only.

The L and M band images show:

i) an unresolved core down to the resolution (FWHM) of
0.24'' (16\,pc) and 0.33'' (22\,pc), respectively at 3.5 and 4.8 \mic. This 
core has already been observed at 3.6 \mic\ by Chelli et al. (1987)
and at 2.2 \mic\ by Marco et al. (1997), Thatte et al. (1997)
  and Rouan et al. 
(1998). The latter give an upper limit of the core size (FWHM) of 0.12'' 
(less than 8 pc).

ii) an elongated structure at P.A. \ap 100\dg\ particularly prominent in the M 
band, but also quite well outlined in the L band. 
This structure is obviously
 coincident with the structure seen in the K band by Rouan et al. (1998) and is
roughly perpendicular to the axis of the inner ionizing cone 
(P.A.=15\dg, Evans 
et al. 1991).
It extends in total over \ap 80\,pc, with a bright spot at each of the  
\ap E\ and \ap W\ edges
at a radius of \ap 25\,pc from the central engine.

iii) an extended emission along the NS direction, almost aligned with the radio 
axis
and the ionizing cone axis. 
At low level isophotes (in particular in the L band), a change in the direction 
of the axis
of this emission can be noticed, reminiscent of a similar change of  direction  of
the radio jet (Gallimore et al. 1996a).

Down to faint levels, the 4.8\,\mic\ thermal infrared emission appears to be
 extended over \ap 3'' in diameter (\ap 210\,pc).

It is striking that the two different AO systems used, ADONIS for the L and M
 bands and PUEO for the K band, reveal a similar structure of the AGN 
environment.
These AO systems are using WFS of different types, Shack-Hartmann for ADONIS
 and curvature for PUEO, as well as deformable mirrors of different types,
 piezo-stack for ADONIS and bimorph for PUEO. The Lucy-Richardson 
 deconvolution 
applied on both data sets uses PSFs obtained in two different ways: we used an
observed stellar PSF in the case of the ADONIS data set -- as the L and M
band data are less sensitive to rapid 
PSF 
fluctuations -- and we used the PSF recovered from the AO loop parameters in the 
case of the
PUEO data set. The two AO experiments differ in many aspects, while leading to a
similar result for the structure of the AGN dusty environment. Therefore we 
are quite confident that this structure is real and not hampered by 
significant AO 
artifacts
(Chapman et al. 1999). Finally it should be noticed that the high resolution 
image of the AGN in \ngc\ obtained in the K band with the AO system at the Keck
 telescope {\it (www2.keck.hawaii.edu/realpublic/ao/ngc1068.html)} 
reveals a comparable structure, pending that a
 precise orientation and a scale be provided for the Keck AO data set.

\subsection{Location of the emission peaks at 3.5 and 4.8\,\mic\ and nature
 of the unresolved core}

As it was not possible to observe simultaneously in the visible and in the 
infrared, 
we took advantage of a characteristic feature of AO
systems which is to preserve
the optical center for all objects: the infrared camera field has a
position fixed in regard to the centroid of the visible counterpart of
the object observed.  
Indeed, by observing a star (PSF or photometric standard), 
we determine a reference position in the infrared image
to within the precision we are aiming at in this study 
(better than one infrared camera pixel, 0.1'').
Any offset of
the galaxy infrared peak relatively to the star infrared peak would then
reveal an intrinsic offset between the galaxy infrared light peak and the galaxy
visible light peak.
This is a method for positioning  infrared 
versus visible sources in the AGN.

To improve the precision, we have fitted the PSF and the \ngc\ emission peaks by  
Gaussian profiles. The L and M band peaks in \ngc\ are coincident within the 
positional precision given above.
We have also derived the position of this L and M peak in \ngc\ with respect to 
the 
visible peak, following the procedure described in Sect. 3: it 
is offset by $0.3\pm 0.05$''  
S and $0.1\pm 0.05$'' W of the visible continuum peak and therefore is 
found to be coincident with the K band emission peak (Marco et al. 1997),
 within the error bars.

\begin{table}[t]
\caption{Photometric data for \ngc. Observed fluxes are in Jy.}
\begin{center}
\begin{tabular}{lcccccl}
\hline
filter  &\multicolumn{6}{c}{aperture radius} 	\cr
 	 &0.3''	&0.4''	&0.5''	&0.6''	&0.9''	&1.5'' \cr
\hline
L	 &0.92 &1.34 &1.78 &2.09  &2.76  &3.40  	\cr
M	 &2.27 &3.23 &4.41 &5.30  &7.18  &9.73 	\cr
\hline
\end{tabular}
\end{center}
\label{flux}
\end{table}

Therefore, the compact core at 3.5 and 4.8 \mic\ can be identified with
the unresolved core detected at 2.2\mic\
 (Marco et al. 1997; Thatte et al. 1997; Rouan et al. 1998), itself found
 to be coincident with 
 the mid-infrared emission peak at 12.4\mic\ (Braatz et al. 1993), 
 the radio source S1 (Gallimore et al. 1996a, b)
and the center of symmetry of the UV polarization map (Capetti et al. 1995).
This strengthens considerably the interpretation 
of the core infrared emission originating from hot/warm dust in the 
immediate surroundings of the central engine.

\subsection{The torus-like structure at 3.5 and 4.8 \mic }

The location, position angle and extension of the P.A. \ap 100\dg\ structure 
are strongly suggestive of a dusty/molecular torus. 
The two bright spots on the edges of the structure 
outline the ``disky'' nature of the torus, up to a radius of 
\ap 40\,pc  
from the central engine.
This dusty/molecular torus would be responsible for the collimation of 
 UV radiation from the central
engine, leading to the ionizing cone (Pogge 1988; Evans et al. 1991).
The overall spatial extension of the torus is found to be \ap 80\,pc 
at 3.5 and 
4.8\mic, while 
it appears to be slightly smaller at 2.2\mic, \ap 50\,pc. 
Under the very simple assumption of optically thick grey-body dust radiation
(Barvainis 1987), it is well understood that the emission
at 2.2\mic\ traces hotter dust (T\ap 1300\,K) than the emission at 4.8 \mic\ 
(T\ap 600\,K).
The observed difference in size would then signal the existence of a 
temperature gradient of the grains across the torus.

\subsection{The North South extended emission}

The NS extended emission (overall extent  \ap 3'' down to faint emission 
levels)
is also detected at 2.2\mic\ on a similar scale (Rouan et al. 1998)
and at 10 and 20 \mic\ on a larger scale, although along a similar P.A.
 (Alloin et al. 1999) .
This structure can be related to the emission of hot/warm dust associated 
with NLR clouds
 identified in the northern side of the ionization cone from HST data 
 (Evans et 
al. 1991) and hidden behind the disc of the galaxy in its southern 
side.
Additional local heating processes, e.g. related to shocks induced by the 
jet propagation , might
be at work as well along the NS extension. The latter suggestion stems from 
the conspicuous change of direction of the emission at 3.5\,\mic , following 
that of the radio jet. Indeed,
Kriss et al. (1992) have shown through the analysis of line emission 
that in \ngc\ the emitting gas in the NLR is partly excited through
shocks triggered by the radio jet.

\section{Fluxes, SED and variability}

The spectral energy distribution (SED)
of the central region of the AGN
is an essential parameter in the modeling.  
To derive this quantity, spatial resolution is obviously 
needed to disentangle the different sources
of emission -- dust, stars, non-thermal source.. -- and, in that respect, 
AO 
observations bring precious informations. 
 
Fluxes at 3.5 and 4.8\,\mic\ have been measured through circular apertures 
centered on the near-infrared peak, 
with a radius varying
from 0.3'' to 1.5'' (22 to 100\,pc). They are depicted in Table
\ref{flux}.
A weak PAH line emission at 3.3\mic\ 
has been detected as well, although no flux calibration is 
available for this observation, unfortunately .

The aperture flux density as a function of radius,
over the region 22 pc $\leq r \leq$ 100 pc,
can be fitted with a power law:
we find $F_L \propto r^{-1.05}$ and $F_M \propto r^{-1.00}$
(where the flux unit is Jy/arcsec$^2$).

\begin{figure}[h]
\resizebox{\hsize}{!}{\includegraphics{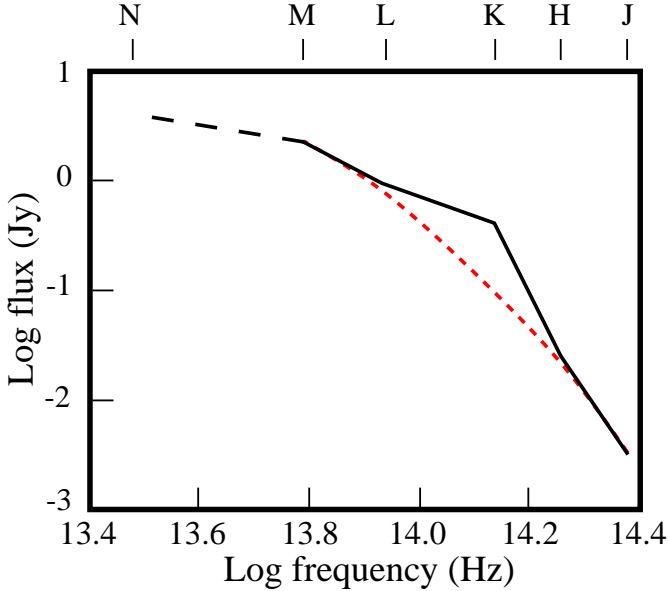}}
\caption{SED of the central 0.6" diameter core (the value for the N band is 
given for 0.8", hence the dashed line between the M and N bands). The dotted 
line represents an interpolation for establishing the lower-limit of the 
dust contribution 
in the K band.} 
\label{SED}
\end{figure}

From the set of AO images (this paper and Rouan et al. 1998), 
as well as high resolution images obtained at 10\,\mic\ (Alloin et al. 1999),
we have reconstructed the SED of the core emission through an 0.6''
diameter diaphragm, as shown in Fig. \ref{SED}. Therefore, this SED
refers only to the central engine and its immediate environment (including 
the inner parts of the dusty/molecular torus as well as some
contribution from the NS extended structure).
In this plot, 
the contribution from the stellar component has been removed in the J 
and H bands, following the spatial profile analysis performed by Rouan et 
al. (1998), while this is not the case in the K band. The part of 
the near-infrared to mid-infrared emission which is arising from hot to 
warm dust grains can be represented by a series of grey-bodies of 
different temperatures and this summation is expected to result in a smooth 
distribution. Yet, an emission bump at 2.2\,\mic\, is observed, which
could be interpreted as the unremoved stellar contribution 
in the K band, within the 0.6'' diameter diaphragm. Under this 
assumption, an upper limit of 75\% for the stellar contribution to the flux at 
2.2\,\mic,
can be derived. This upper limit remains far larger than the 6\% effectively
derived 
by Thatte et 
al. (1997) within a 1'' diameter diaphragm, on the basis of the dilution 
of the equivalent width of a CO absorption feature arising from cold stars.

In any case, a revision of the AGN 
modeling for
\ngc\ should incorporate the 1 to 10 \mic\, SED derived 
for the innermost region 
around the central engine and therefore less affected by dilution from other 
surrounding components (in particular cleaned from the stellar 
contribution).

It is interesting as well to compare these 1996/1997 measurements 
in the infrared to those 
performed  by Rieke \& Low as early as 1975. Therefore, we have derived, for 
their 3'' diameter diaphragm, the 1996/1997 observed fluxes at 1.25, 1.65 
(both 
uncorrected for the stellar contribution), 2.2, 3.4, 4.8 and 10\,\mic, 
from Rouan et al. (1998), from 
the current data set and from  Alloin et al. 
(1999). 

Before examining the temporal behavior of the near-infrared emission in
\ngc, the 1997 flux measurements in the 
K band can be compared to previous determinations. We consider again a 3'' 
diameter diaphragm 
and compare the 1997 AO measurement to classical photometric measurements.
 Generally those are performed through 
much larger diaphragms. However Penston et al. (1974) provide a full set of 
measurements through diaphragms ranging from 12'' to 2'' diameter. 
Leaving aside 
the measurements from this paper which have been flagged down for bad 
transparency or other flaws, we obtain the K magnitude offset when one moves
 from a 12'' diaphragm to a 3'' diaphragm, 
$\Delta$K = 0.85. This magnitude offset is not expected to vary with 
time because it refers to the outer parts of the AGN 
($r >$ 110 pc). Then, from the 12'' diameter diaphragm 
measurements by Glass (1995), who analyzed the variability properties of \ngc ,
we can infer/extrapolate the K magnitude at the date of the AO measurement by 
Rouan et al. 
(1998):  K $=$ 6.81. Applying the magnitude offset computed above between the
 12'' and 3'' diameter diaphragms, we predict that the K magnitude should be
of 7.66 at the date of the Rouan et al. (1998) observation, while the K magnitude 
measured is of
7.26. This agreement is quite satisfactory, given the rather large error-bars
involved in the Penston et al. (1974) data set which was obtained more than 25 
years ago.

A comparison of the fluxes within a 3'' diameter diaphragm at both epochs,
1975 (Rieke \& Low 1975) and 1997 (this paper) is depicted in 
Table \ref{variab}. One notices immediately that a flux increase has occurred 
over this 
time interval: by a factor 2 between 2.2 to 4.8\,\mic, and by 
a factor 1.2, at 10\,\mic. According to the independent photometric monitoring 
by
Glass (1997), the L band (3.5\,\mic) flux
has doubled in 18 years, from 1974 to 1992: our result is in very good 
agreement with his finding.
 
Because of the possibly complex way through which the UV-optical photons 
illuminate and heat the dust grains (direct and/or indirect illumination
via scattering on the mirror to the N of the central engine or on
dusty regions further out, Miller et al. 1991), it is not possible 
to infer the size of the dust component from light-echo 
effects acting 
differentially in the near-infrared and mid-infrared bands.

\begin{table}
\caption{Comparison of the infrared fluxes (given in Jy) at two epochs  in a 3''
diameter diaphragm}
\begin{center}
\begin{tabular}{lcccccc}
\hline
        &J      &H      &K      &L      &M      &N      \\
1997    &0.13   &0.3    &0.8    &3.4    &9.7    &21.5   \\
1975    &       &       &0.3    &1.7    &5.3    &17.9   \\
\hline
\end{tabular}
\end{center}
\label{variab}
\end{table}

\section{Colors, dust temperature and mass of dust}

Following
Barvainis (1987), we assume the emissivity of the grains to 
depend on wavelength as $\lambda ^{-1.6}$ and we
use a simple model of optically thick grey-body dust emission:
${\rm \nu}^{1.6} B_{\rm \nu}(T_{\rm gr})$.

In the general situation of an AGN, 
the grain temperature varies strongly with distance to the central engine,
following a power law (Barvainis 1987):
$T_{\rm gr}(r) = 1650 L_{UV,46}^{0.18} r^{-0.36}$ K, where
$L_{UV,46}$ is the UV luminosity in units of $10^{46}$ ergs\,s$^{-1}$ 
and $r$ the radial distance in parsecs.

\subsection{Color gradients in the near-infrared}

\subsubsection{Colors of the core}
Owing to our limitation in spatial resolution in the M band (FWHM = 0.33''),
 we can measure at best the [L-M] and [K-L] colors of the core through an 
0.6'' diameter diaphragm centered on the near-infrared emission peak. We find
[L-M] = 1.6 $\pm$ 0.4 and [K-L] = 1.8 $\pm$ 0.2. It must be noticed however 
that the contribution of stellar light in the K band (Thatte et al. 1997) 
has not been 
removed at this stage and that the observed [K-L] color does not relate only 
to the dust emission. 
\subsubsection{Colors of the extended structure}
 As for the colors of the sources forming the extended structures, and again
 because of different spatial resolutions in the K, L and M bands, we have 
 considered the mean colors in a 
ring which extends from $r$ = 0.3'' to $r$ = 0.5''. This ring does 
include the 
emitting regions forming the extended structures 
along the two directions  P.A. \ap 100\dg\ and NS,
 at 3.5 and 4.8\,\mic , but 
excludes in part the secondary peaks which delineate the extended
 structures at 2.2\,\mic . Therefore, at 2.2\,\mic , the ring includes more 
 of the diffuse contribution possibly related with the stellar core analyzed
 by Thatte et al. (1997). The colors found for the ring, representative of 
 a mean 0.4'' radius, 
are [L-M] = 1.6 $\pm$ 0.4 and [K-L] = 2.8 $\pm$ 0.2.

\subsection{Dust temperature}

The advantage of interpreting the [L-M] color is that the L and M band flux 
contributions
are known to arise almost entirely from the dust component. 
Within the limitation in spatial resolution from the M band data set, we do 
not detect any [L-M] color gradient within the central 1'' region of 
\ngc. 

 Under the simple assumption of optically thick grey-body emission from the
 dust component,
the observed [L-M] color corresponds to a grain temperature 
$T_{gr}$\ap 480\,K. 
The foreground extinction to the core has been calculated by several authors
(Bailey et al. 1988; Bridger et al. 1994; Efstathiou et al. 1995; Young et al. 
1995; Veilleux et al. 1997;
Glass 1997; Thatte et al. 1997; Rouan et al. 1998) 
and an estimate of $A_v$\ap 30 mag is retained. 
Applying a correction for such an extinction, 
 we deduce an intrinsic color [L-M] = 0.8 $\pm$ 0.4 and 
 $T_{gr}$\ap 700\,K. The 
absolute luminosity $L_{UV,opt}$ of the AGN in \ngc\, -- assumed here to be 
the unique heating source of the dust grains in the AGN environment -- can 
be approached
only indirectly and is still quite uncertain. From the analysis by 
Pier et al. (1994) who have examined  
various methods for  deriving the absolute luminosity of the AGN in \ngc\, 
and summarize the current knowledge on this question, we deduce
 L$_{UV,opt}$ = 4 $10^{44}$ erg s$^{-1}$. However it should be noted that 
 this figure has been obtained 
assuming a 
reflected light fraction f$_{refl}$ \ap 0.01, while the consideration of 
ionized gas in 
regions further out inside the ionization cone (ENLR) leads to a value 
f$_{refl}$ \ap 0.001
(Bland-Hawthorn et al. 1991). 
Then a value as high as L$_{UV,opt}$ = 4 $10^{45}$ erg s$^{-1}$ should 
be envisaged as well. 
In addition, most of these estimates have been derived without taking 
into account the 
fraction of UV-optical flux which provides the dust grains heating: 
already the K magnitude of the innermost core (FWHM = 0.12''), 9.3,
corresponds to an energy output of \ap 8 $10^{41}$ erg s$^{-1}$. It might
 be important to consider the energy 
radiated in the near- to mid-infrared bands for 
the evaluation of L$_{UV,opt}$. In conclusion, the figures currently 
available for L$_{UV,opt}$ in \ngc\ might be lower limits.

Still, under the simple model of
optically thick grey-body dust emission, reaching  
 $T_{gr}$\ap 480\,K at $r$ = 28 pc requires  L$_{UV,opt}$ = 8 $10^{45}$ erg 
s$^{-1}$, a value roughly consistent with the highest figure given above for
L$_{UV,opt}$ in \ngc . This figure goes up to  
L$_{UV,opt}$ = 6.5 $10^{46}$ erg s$^{-1}$
if the grain temperature is of 700\,K at $r$ = 28 pc (extinction-corrected 
estimate), pushing \ngc\ to the 
limit
between AGN and quasars. This question certainly deserves further attention
and above all the consideration of a more elaborated model of  
the dust region, with regard to its geometry and heating. This is 
beyond the scope of the current
paper and will be discussed in the future.

The [K-L] color in the extended structures can be contaminated by some stellar 
contribution in the
2.2\,\mic\ band. From grains at $T_{gr}$\ap 480\,K, which are the dominant 
contributors, we expect a [K-L]$_{dust}$ 
color of 4.0. Given the observed [K-L] value, we deduce that the 
pourcentage of the flux at 2.2\,\mic\ which arises from the dust component 
(with at most $T_{gr}$\ap 480\,K) is of 30\%. For this estimate, we have 
not considered
any correction for extinction.

How can the lack of [L-M] color gradient between the 0.6'' diameter core and 
the extended structures be explained? Given the unresolved and intense 
core emission at 
2.2\,\mic\ (FWHM = 0.12'' from Rouan et al. 1998) it can be inferred that 
the hottest dust grains are extremely 
confined and located at a radius less than 4 pc. With 
 L$_{UV,opt}$ = 8 $10^{45}$ erg s$^{-1}$, they would be present only up 
 to $r$ = 1.1 pc. Hence, there must exist a very steep dust temperature 
 gradient close to the 
central heating source. Such a few parsec scale corresponds to a resolution 
which is well beyond that accessible
at 3.5 \& 4.8 \mic . In fact the L and M emission we are measuring in 
an 0.6'' diameter 
core is already strongly dominated by the 
warm grains. Because of this suspected steep temperature gradient, the 
procedure applied previously to derive the stellar contribution in the 
0.3'' to 0.5'' radius ring cannot be used in the core.

\subsection{Mass of the hot dust}

The mass of dust associated with the near-infrared emission can be
estimated only in a rough way, as it depends on the (unknown) 
grain composition and grain size distribution.
Assuming graphite grains and following Barvainis (1987),
the infrared spectral luminosity of an individual
graphite grain is given by:
$L^{\rm gr}_{\rm \nu,ir}=4\pi a^2 \pi Q_{\rm \nu} B_{\rm \nu}(T_{\rm gr}){\rm 
~ergs ~s^{-1} ~Hz^{-1}}$
where $a$ is the grain radius, $Q_{\rm \nu}=q_{\rm ir} {\rm ~\nu
^{\gamma}}$ 
is the absorption efficiency of the grains, and $B_{\rm
\nu}(T_{\rm gr})$ is the Planck function for a grain temperature $T_{\rm
gr}$. Following Barvainis (1987), we take $a$=0.05\,\mic\, 
and in the near-infrared, $q_{\rm ir}=1.4 \x 10^{-24}$, $\gamma =1.6$
leading to $Q_{\rm \nu}$=0.058 (for the K band).

Because no [L-M] color gradient is detected towards the 0.6'' diameter core,
we consider the simple case of 2 populations of dust grains in that region,
hot grains at T\,=\,1500\,K and warm grains at T\,=\,500\,K, 
matching the extreme values in that region. 
Solving the equation 
$F_{\rm \nu , measured}= x F_{\rm \nu , 1500\,K} + y F_{\rm \nu , 500\,K}$
for the three bands available, K, L and M, one derives 2\,10$^{45}$ and 
9\,10$^{47}$, for
the number of grains at 1500\,K and 500\,K  respectively, in the 0.6'' 
diameter core.
This indicates that there are \ap 450 times more warm dust grains than hot 
dust grains
in the 0.6'' diameter core. The warm dust grains dominate the [L-M] color.
With a grain density $\rho$=2.26~g\,cm$^{-3}$, the mass of warm 
dust grains is found to be
 $M$(warm dust)\ap 0.5~$M_{\odot}$.
This mass is above that of hot dust grains detected in two
Seyfert 1 nuclei: 0.05~$M_{\odot}$ in the case of {\sc ngc}\,7469 (Marco et al. 
1998)
and 0.02~$M_{\odot}$ in the case of Fairall 9  (Clavel et al. 1989). This 
result supports the fact that only a small fraction of the dust present in 
the torus is heated close to its sublimation temperature.

\section{Grain number density in the warm dust}

In \ngc, like in other AGN (see for instance {\sc ngc}\,7469),
 dust is present also in the NLR region (see Fig. \ref{lmimage}).
It has been shown in Sect. 5 that the infrared emission in the L \& M bands
over the region $22 \leq r \leq 100$ pc follows a power-law.
In addition the temperature of the 
grains deduced from the [L-M] color has been found to 
 remain roughly constant with radius.
Thus, we can derive the radial profile of the warm grain number
density in the NLR region, $\eta (r) \propto r^{- \beta}$. 
Under optically thin conditions, probably 
applicable in the NLR, the brightness is directly proportional to the 
grain number
density. Then the observed brightness power-law leads to $\beta$=1.0, 
suggesting a concentrated grain distribution
(a uniform grain density would correspond to $\beta$=0). 
In {\sc ngc}\,7469, the warm dust component led to a figure for $\beta$ 
around 1.5 
(Marco et al. 1998).

\section{Comparison with thick tori model predictions}

Several torus models have been developed so far to explain the
obscuration of the BLR and UV/X-ray continuum sources along some lines
of sight (AGN unification scheme). Some of these models are generic, 
while others have been designed to match the case of \ngc .

Pier \& Krolik (1992a, b, 1993) propose a thick, parsec-scale,
uniform density torus  
illuminated by a nuclear 
source. The dust can be heated up to the effective
temperature of the nuclear radiation at the inner edge of the
torus. They investigate models with effective temperatures between 
500 K and 2000 K. Such a model can explain the unresolved core observed 
at 2.2, 3.5 and 4.8 \mic\ with AO 
in the particular case of \ngc. Does it explain the extended near infrared 
emission also revealed by these observations? Indeed, extended emission
over 1'' to 2'' could result from reflected radiation from the torus and/or
dust in the NLR. Therefore, this model accounts for most of the features 
unveiled with high resolution imaging in the near infrared.

Efstathiou \& Rowan-Robinson (1995) propose a model with a very thick
tapered disk.  They assume
the melting temperature of all  dust grains to be identical (1000\,K), 
but consider a radial distribution of the grain physical parameters (size and 
chemical composition).
In the case of \ngc, Efstathiou et
al. (1995) have shown that the torus emission alone cannot account for the
total infrared emission. They attribute the excess infrared emission
to a distribution of optically thin dust with $\beta$=2
in the NLR region. Their model is in disagreement with the steep grain 
temperature 
gradient across the torus, which we infer to exist close to the central 
engine. Conversely, the dust postulated to be present in the NLR by their model
is indeed detected with the AO data set.

Granato \& Danese (1994) and
Granato et al. (1996, 1997) developed a simple thick
($\tau_e>$30) torus model extended over several hundreds pc. To
minimize the number of free model parameters, they have
adopted a dust density distribution constant with radial
distance from the nuclear source. But,
they do not rule out the possibility, in the case of smaller values of 
the
optical depth ($\tau _e$=1.5), that a more concentrated density
distribution exist. Their predicted size and shape for the near infrared 
emission are compatible with those derived through AO observations at 
2.2, 3.5 and 4.8 \mic.

A revised modeling of the AGN in \ngc\ is timely, owing to the emergence of
sub-arcsec resolution images in the near infrared (AO techniques) and in 
the millimeter range (interferometric techniques), giving direct access to 
the dust 
and molecular environment of the central engine.

\section{Conclusion}

The observation, for the first time at high angular resolution, of 
\ngc\ at 3.5 and 4.8 \mic\ provides new informations to build a more 
realistic model of this AGN under the current unification scheme.\\
As regard to the AGN structure, we do observe: \\
(i) an unresolved core, already known at 2.2\,\mic\ to have a size (FWHM) less 
than 
8\,pc, and interpreted as the inner region of a dusty/molecular torus in 
which the central engine of \ngc\ is embedded, \\
(ii) along P.A.\ap 100\dg, an extended emission up to 40\,pc 
on either side of the 
core, particularly prominent at 4.8\,\mic. This structure, also detected
at 2.2\,\mic\ up to 20\,pc on either side of the core, is coincident in P.A.
with the parsec-scale disc of ionized gas detected with VLBI (P.A.\ap 110\dg ), 
and is found to
be roughly perpendicular to the axis of the ionization cone in \ngc . 
We interpret it as the trace, up to a 40\,pc radius, of this dusty/molecular
torus seen edge-on,\\
(iii) an extended emission along the NS direction up to 50\,pc from the core
and with rather symmetrical properties on either side, both at 3.5 and 4.8 
\mic . Again, this extended emission is detected , both 
at 2.2\,\mic\ on a similar scale, and at 10 and 20 \mic\ on a slightly 
larger scale. It reveals 
the presence of dust in the NLR, heated both by hard 
radiation within the ionizing cone and by shocks associated with 
the AGN radio jet.\\
As regard to the dust temperature and dust distribution in the central arcsec
of the nucleus, we get a final picture as follows. As close as $r$ \ap 1\,pc 
from the
central engine, the dust is heated up to its evaporation temperature, 
1500\,K. Then the
dust temperature declines very rapidly and reaches  T\ap 500\,K at 
$r <$ 28\,pc.
 The total mass of warm dust within a 22\,pc radius region
 is found to be around 0.5\,$M_{\odot}$.
 
 It is observed as well 
 that the near infrared flux of \ngc , in the 2.2 to 4.8 \mic\ range,
has increased by a factor two over some 20 years, while in the 10 \mic\
window the flux increase is only by a factor 1.2. 

Our results bring observational evidence of a dusty
torus in the AGN of \ngc. They further support AGN modelling in
the framework of the unification scheme: a thick torus 
surrounding a central engine (black hole and accretion disc). Although 
several models of the AGN in \ngc\ are available, none
matches in detail all the aspects of the current  
near-infrared results obtained at a subarsec scale.
Such new observational constraints make it both timely 
and exciting to run updated models. Yet, we are aware that the most 
convincing and undisputable argument for the presence of the torus-like 
structure will come from a study of the kinematics of the gas within 
the 100 central 
parsec 
of \ngc . We expect such information to be soon obtained from ISAAC/ANTU
observations on Paranal.

\acknowledgements{We warmly thank J.P. Veran and E. Gendron for useful 
discussions, 
F. Lacombe for his help on the data reduction and Z. Tsetanov for his precious
contribution in deriving the
HST composite image seen by the WFAS. We acknowledge as well precious comments 
from an anonymous referee.}

\end{document}